\documentclass[amsmath,amssymb,showpacs,preprint]{revtex4}

\usepackage{epsfig}
\usepackage{graphicx}
\usepackage{xcolor}

\begin{document}

\title{Finslerian dipolar modulation of the CMB power spectra at scales $2<l<600$}

\author{Xin Li $^{1,2}$}
\email{lixin1981@cqu.edu.cn}
\author{Hai-Nan Lin $^1$}
\email{linhn@ihep.ac.cn}
\affiliation{$^1$Department of Physics, Chongqing University, Chongqing 401331, China\\
$^2$CAS Key Laboratory of Theoretical Physics, Institute of Theoretical Physics, Chinese
Academy of Sciences, Beijing 100190, China}

\begin{abstract}
A common explanation for the CMB power asymmetry is to introduce a dipolar modulation at the stage of inflation, where the primordial power spectrum is spatially varying. If the universe in the stage of inflation is Finslerian, and if the Finsler spacetime is non-reversible under parity flip, $x\rightarrow-x$, then a three dimensional spectrum which is the function of wave vector and direction is valid. In this paper, a three dimensional primordial power spectrum with preferred direction is derived in the framework of Finsler spacetime. It is found that the amplitude of dipolar modulation is related to the Finslerian parameter, which in turn is a function of wave vector. The angular correlation coefficients are presented, and the numerical results for the anisotropic correlation coefficients over the multipole range $2<l<600$ are given.
\end{abstract}

\maketitle

\section{Introduction}

The cosmological principle states that the space is homogenous and isotropic on large scales. It is the foundation of the standard cosmological model, i.e., the $\Lambda$CDM model\cite{Sahni,Padmanabhan}. Although the primordial energy density perturbation generated by the vacuum fluctuation of the inflation field \cite{Starobinsky,Sato:1981,Guth:1981,Linde:1982,Albrecht:1982} will cause small perturbative cosmic anisotropy, most inflation models \cite{Riotto} still preserve the homogeneity and isotropy of the cosmic microwave background (CMB) \cite{Ade:2013sjv,Adam:2015rua}. However, the recent released data of Planck satellite show that the CMB temperature map possesses power asymmetry \cite{Power asymmetry1,Power asymmetry2,Akrami:2014eta}, which has also been observed independently by WMAP satellite \cite{Eriksen:2003db,Hansen:2004vq,Eriksen:2007pc,Hansen:2008ym,Hoftuft:2009ym}. This implies that the statistical isotropy of CMB may be violated.

The power asymmetry can be described as a dipolar modulation of the isotropic power \cite{Gordon}, i.e., $\Delta T(\hat{n})=\Delta T_{\rm iso}(\hat{n})(1+A\hat{p}\cdot\hat{n})$, where $\Delta T_{\rm iso}$ denotes the statistically isotropic temperature map, $\hat{n}$ is the direction of the temperature fluctuation, $\hat{p}$ is the preferred direction of the universe, and $A$ is the dipole amplitude. One possible explanation for the power asymmetry is the modulation of the primordial power spectrum, i.e., $P(k,\mathbf{r})=P(k)(1+2A\hat{p}\cdot\mathbf{r}/r_{ls})$ \cite{Dai:2013,Lyth:2013}, where $r_{ls}$ is the distance to the last scattering surface. One should notice that the primordial power spectrum $P(k,\mathbf{r})$ is a spatially-varying power spectrum, but not a three dimensional spectrum $P(\mathbf{k})$ with preferred direction \cite{Pullen}. This is due to the fact that the background space for Fourier transformation in the $\Lambda$CDM model is the Euclidean space, while the Euclidean space is reversible under parity flip, $\mathbf{x}\rightarrow-\mathbf{x}$. This property guarantees that the Fourier component $\delta(\mathbf{k})$ satisfies the relation $\delta(-\mathbf{k})=\delta^\ast(\mathbf{k})$.

The violation of statistical isotropy of CMB implies the violation of rotational symmetry. Several anisotropic inflationary models have been built to solve the power asymmetry problem \cite{Erickcek:2008sm,Lyth:2013vha,McDonald:2013qca,Jazayeri:2014nya,Assadullahi:2014pya,Kenton:2015jga,Byrnes:2015dub,Byrnes:2016uqw}. Usually, the background spacetime of anisotropic inflation model is described by Bianchi spacetime \cite{Bianchi spacetime}. However, the widely discussed Bianchi type I spacetime \cite{Koivisto:2008,Koivisto:2011} is reversible under parity flip, $\mathbf{x}\rightarrow-\mathbf{x}$. It means that inflationary models based on Bianchi type I spacetime could not generate primordial dipolar modulation. Randers space\cite{Randers}, as a special type of Finsler space \cite{Book by Bao}, is non-reversible under parity flip. In Ref.\cite{Finsler dipolar modulation}, we have studied anisotropic inflation in which the background spacetime is taken to be Finslerian. This Finslerian background spacetime breaks rotational symmetry and induces parity violation. The dipolar modulation of the primordial power spectrum is given in Ref.\cite{Finsler dipolar modulation}. At large scales ($l<100$), this model could approximately explain the released data of Planck satellite for the power asymmetry \cite{Power asymmetry1,Power asymmetry2}. However, the amplitude of the dipole modulation in this model is a constant, which is contradicted to the fact that the amplitude of the dipole modulation is scale-dependent at intermediate scales $100<l<600$ \cite{Hanson,Akrami:2014eta,Ade:2013nlj,Flender:2013jja,Ade:2015hxq}. Recently, Ref.\cite{Aiola} used the CMB temperature maps released by Planck satellite to constrain the scale dependence of the amplitude of the dipole modulation over the multipole range $2<l<600$. In this paper, we try to build a inflation model in Finsler spacetime such that it generate the dipolar modulation of the primordial power spectrum, and at the same time the amplitude of the dipolar modulation is scale-dependent.

The rest of the paper is arranged as follows: In Section \ref{sec:anisotropy}, we briefly introduce the basic concepts of Finsler spacetime, and derive the primordial power spectrum for the quantum fluctuation of inflation field in Finsler spacetime. In Section \ref{sec:field eq}, we derive the gravitational field equations in the perturbed Finslerian background spacetime, and obtain a conserved quantity outside the Hubble horizon. In section \ref{sec:spectrum}, we derive the angular correlation coefficients in our anisotropic inflation model, and plot the numerical results of the angular correlation coefficients that describes the anisotropic effect. Finally, conclusions and remarks are given in Section \ref{sec:conclusion}.

\section{Inflationary field in Finsler spacetime}\label{sec:anisotropy}
Finsler geometry \cite{Book by Bao} is a generalization of Riemann geometry  without quadratic restriction on the metric. The basic quantity in Finsler geometry is the Finsler structure $F$, which is defined on the tangent bundle of a manifold $M$, with the property $F(x,\lambda y)=\lambda F(x,y)$ for any $\lambda>0$, where $x\in M$ represents position and $y$ represents velocity. The second order derivative of the squared Finsler structure with respect to $y$ gives the Finslerian metric \cite{Book by Bao}
\begin{equation}
g_{\mu\nu}\equiv\frac{\partial}{\partial
y^\mu}\frac{\partial}{\partial y^\nu}\left(\frac{1}{2}F^2\right).
\end{equation}
Hereafter, Greek indices are lowered and raised by $g_{\mu\nu}$ and its inverse matrix $g^{\mu\nu}$, respectively. The Finslerian metric reduces to Riemannian metric if and only if $F^2$ is quadratic in $y$. The Finslerian spacetime is fully described by the Finsler structure $F$.

In this paper, we propose that the background spacetime of the universe is the Randers spacetime during the stage of inflation. The Randers spacetime is a special kind of Finsler spacetime which have been widely discussed elsewhere \cite{Chang:2013lfm,Chang:2014ndh,Chang:2014smk,Chang:2015nbc,Finsler dipolar modulation}. The Finsler structure of Randers spacetime takes the form
\begin{equation}\label{FRW like}
F^2=y^ty^t-a^2(t)F_{Ra}^2,
\end{equation}
where $F_{Ra}$ is the Randers space \cite{Randers}
\begin{equation}
F_{Ra}=\sqrt{\delta_{ij}y^iy^j}+\mathbf{b}(\mathbf{x})\cdot \mathbf{y},~~\mathbf{b}\cdot \mathbf{y}\equiv\delta_{ij}b^iy^j.
\end{equation}

Now, we focus on finding the equation of motion for the inflationary field. The action for the single-field ``slow-roll'' scalar field is given by
\begin{equation}\label{action}
S=\int d^4x\sqrt{-g}\left(\frac{1}{2}g^{\mu\nu}\partial_\mu\phi\partial_\nu\phi-V(\phi)\right).
\end{equation}
Noticing that the determinant of the background spacetime (\ref{FRW like}) is given by $g=-a^3F_{Ra}^2/(\delta_{ij}y^iy^j)$, we find from the action (\ref{action}) that the equation of motion for the quantum fluctuation \cite{Rev inflaton1,Rev inflaton2} field $\delta\phi$ is given by
\begin{equation}\label{EOM}
\delta\ddot{\phi}+3H\delta\dot{\phi}-a^{-2}\bar{g}^{ij}\left(\partial_i\partial_j\delta\phi+2\partial_i\delta\phi\frac{\partial(\mathbf{b}\cdot \mathbf{y})}{F_{Ra}}\right)-a^{-2}\partial_i\delta\phi\partial_j\bar{g}^{ij}=0,
\end{equation}
where the dot denotes the derivative with respect to time, $H\equiv\dot{a}/a$, and $\bar{g}^{ij}$ is the Finslerian metric on Randers space $F_{Ra}$. Throughout this paper, the indices labeled by Latin letter are lowered and raised by $\bar{g}_{ij}$ and its inverse matrix $\bar{g}^{ij}$, respectively. In Finslerian gravity, the direction of $\mathbf{y}$ will appear in the field equation and affect the gravitational field. During the stage of inflation, following our former treatment \cite{Finsler dipolar modulation,Finsler tensor spectrum}, we take the direction of $\mathbf{y}$ to be parallel to the wave vector $\mathbf{k}$ in the momentum space. Therefore, in the momentum space, equation (\ref{EOM}) can be simplified to
\begin{equation}\label{EOM1}
\delta\ddot{\phi}+3H\delta\dot{\phi}-a^{-2}k^2(1+5(\mathbf{b}\cdot\hat{\mathbf{k}}))\delta\phi=0,
\end{equation}
where $\hat{\mathbf{k}}$ denote the unit vector along $\mathbf{k}$. We have neglected the $b^2$ term in deriving equation (\ref{EOM1}).
Then, following the standard quantization process in the inflation model \cite{Mukhanov book}, we can obtain the primordial power spectrum from the solution of equation (\ref{EOM1}). It is of the form
\begin{equation}\label{spectrum delta phi}
\mathcal{P}_{\delta\phi}(\mathbf{k})\simeq\mathcal{P}_0(k)\left(1-\frac{15}{2}\mathbf{b}\cdot\hat{\mathbf{k}}\right),
\end{equation}
where $\mathcal{P}_0$ is the isotropic power spectrum for $\delta\phi$, which depends only on the magnitude of wave vector $\mathbf{k}$.

\section{The conserved quantity at horizon crossing}\label{sec:field eq}

In the standard inflation model \cite{Mukhanov book}, the primordial power spectrum that links to observed quantities of CMB is the spectrum of the comoving curvature perturbation, which is conserved outside the Hubble horizon. In this section, we focus on finding a counterpart to the comoving curvature perturbation in Finsler spacetime. In standard process \cite{Mukhanov book}, the conserved quantity at horizon crossing, i.e., the comoving curvature perturbation is derived by using gravitational field equations and the long wavelength approximation, namely, we ignore the terms that are proportional to $k^2$. This approximation is valid since we are interested in the modes of quantum fluctuations that are outside the Hubble horizon during the stage of inflation. In Finslerian inflation, one should notice that the Finslerian parameter $\mathbf{b}$ is only a function of spatial coordinates, and the long wavelength approximation means that the second order derivative of $\mathbf{b}$ can be neglected. Thus, we can expect from the qualitative analysis that the conserved quantity at horizon crossing in Finslerian inflation is the same to the comoving curvature perturbation in standard inflation model.

In this paper, we just consider the scalar perturbation of Finsler structure (\ref{FRW like}). It is of the form
\begin{equation}\label{FRW like perturbed}
F^2=(1+2\Psi(t,\mathbf{x}))y^ty^t-a^2(t)(1+2\Phi(t,\mathbf{x}))F_{Ra}^2,
\end{equation}
where $\Psi$ and $\Phi$ are scalar perturbations. There are two important geometric quantities in Finslerian gravity. One is the geodesic spray coefficients $G^\mu$, which is related to the first order variation of Finslerian length. The first order variation of Finslerian length gives the geodesic equation, which reads
\begin{equation}
\label{geodesic}
\frac{d^2x^\mu}{d\tau^2}+2G^\mu=0,
\end{equation}
where
\begin{equation}
\label{geodesic spray}
G^\mu=\frac{1}{4}g^{\mu\nu}\left(\frac{\partial^2 F^2}{\partial x^\lambda \partial y^\nu}y^\lambda-\frac{\partial F^2}{\partial x^\nu}\right).
\end{equation}
It can be proven that the Finsler structure $F$ is a constant along the geodesic \cite{Book by Bao}. Substituting the Finsler structure (\ref{FRW like perturbed}) into equation (\ref{geodesic spray}), we obtain
\begin{eqnarray}\label{geodesic spray t}
G^t&=&\frac{1}{2}\dot{\Psi}y^ty^t+\Psi_{,i}y^ty^i+\frac{1}{2}\left[a\dot{a}(1+2\Phi-2\Psi)+a^2\dot{\Phi}\right]F^2_{Ra},\\ \label{geodesic spray i}
G^i&=&\left(H+\dot{\Phi}\right)y^ty^i+\Phi_{,k}y^ky^i+\frac{1}{2a^2}\Psi_{,j}\bar{g}^{ij}y^ty^t-\frac{1}{2}\bar{g}^{ij}\Phi_{,j}F^2_{Ra}\nonumber\\
   &&+\frac{1}{2}\bar{g}^{ij}\bar{g}_{jk,m}y^my^k-\frac{1}{2}\bar{g}^{ij}F_{Ra}(\mathbf{b}\cdot \mathbf{y})_{,j},
\end{eqnarray}
where the comma denotes the derivative with respect to spatial coordinate $x$. Another important geometric quantities is the Ricci scalar, which is related to the second order variation of Finslerian length. In physics, the second order variation of Finslerian length gives the geodesic deviation equation which describes the gravitational effect between two particle moving along the geodesics. The analogy between geodesic deviation equations in Finsler spacetime and Riemannian spacetime gives the vacuum field equation in Finsler gravity \cite{Finsler bullet,Finsler BH}, namely, the vanish of the Ricci scalar.

In Finsler geometry, the Ricci scalar is given by \cite{Book by Bao}
\begin{equation}\label{Ricci scalar}
{\rm Ric}\equiv\frac{1}{F^2}\left(2\frac{\partial G^\mu}{\partial x^\mu}-y^\lambda\frac{\partial^2 G^\mu}{\partial x^\lambda\partial y^\mu}+2G^\lambda\frac{\partial^2 G^\mu}{\partial y^\lambda\partial y^\mu}-\frac{\partial G^\mu}{\partial y^\lambda}\frac{\partial G^\lambda}{\partial y^\mu}\right).
\end{equation}
The Ricci scalar is a geometric invariant quantity and is insensitive to the connections. It only depends on the Finsler structure $F$. Substituting equations (\ref{geodesic spray t},\ref{geodesic spray i}) into equation (\ref{Ricci scalar}), we obtain
\begin{eqnarray}
F^2{\rm Ric}&=&-3\frac{\ddot{a}}{a}y^ty^t+(a\ddot{a}+ 2\dot{a}^2)F^2_{Ra}+y^ty^t\left(a^{-2}\bar{g}^{ij}\Psi_{,i,j}- 3\ddot{\Phi}+3H(\dot{\Psi}-2\dot{\Phi})\right)\nonumber\\
&&+y^ty^i\left(4H\Psi_{,i}-4\dot{\Phi}_{,i}\right)-y^iy^j\left(\Psi_{,i,j}+\Phi_{,i,j}\right)\nonumber\\
&&+F_{Ra}^2\left((2\dot{a}^2+a\ddot{a})(2\Phi-2\Psi)+a\dot{a}(6\dot{\Phi}-\dot{\Psi})+a^2\ddot{\Phi}-\bar{g}^{ij}\Phi_{,i,j}\right)\nonumber\\
&&+\Psi_{,i}\bar{g}^{ij}_{~,i}y^ty^t-2F_{Ra}\Phi_i\bar{g}^{ij}(\mathbf{b}\cdot \mathbf{y})_{,j}-F_{Ra}\bar{g}^{ij}(\mathbf{b}\cdot \mathbf{y})_{,i,j}+\frac{F_{Ra}}{2}\bar{g}^{ij}b_{i,j,k}y^k\nonumber\\
\label{Ricci scalar1}
&&+\frac{y^i}{2F_{Ra}}(\mathbf{b}\cdot \mathbf{y})_{,i,j}y^j+\frac{1}{2}\bar{g}_{ij,m,n}y^n(\bar{g}^{jm}y^i-\bar{g}^{ij}y^m),
\end{eqnarray}
where we have omitted the term that is proportional to $\frac{\partial\bar{g}^{ij}}{\partial y^i}$. From the relation $\frac{\partial\bar{g}^{ij}}{\partial y^i} y_j=0$, and noticing that the direction of $\mathbf{y}$ is parallel to the wave vector $\mathbf{k}$ in momentum space, one can infer that the term proportional to $\frac{\partial\bar{g}^{ij}}{\partial y^i}$ in the right-hand-side of equation (\ref{Ricci scalar1}) should be vanishing. In Refs.\cite{Chang:2015nbc,Finsler BH}, it has been proven that the gravitational field equation in Finsler spacetime is of the form
\begin{equation}\label{field equation}
{\rm Ric}^\mu_\nu-\frac{1}{2}\delta^\mu_\nu S=8\pi G T^\mu_\nu,
\end{equation}
where $T^\mu_\nu$ is the energy-momentum tensor. Here the Ricci tensor is defined by \cite{Akbar}
\begin{equation}\label{Ricci tensor}
{\rm Ric}_{\mu\nu}=\frac{\partial^2\left(\frac{1}{2}F^2 {\rm Ric}\right)}{\partial y^\mu\partial y^\nu},
\end{equation}
and the scalar curvature in Finsler spacetime is given by $S=g^{\mu\nu}{\rm Ric}_{\mu\nu}$.
Combining equations (\ref{Ricci scalar1}--\ref{Ricci tensor}), we obtain the background field equations
\begin{eqnarray}\label{field eq t}
3H^2&=&8\pi GT^0_{(0)0},\\
\label{field eq i}
\frac{2\ddot{a}}{a}+H^2&=&-\frac{8\pi}{3} G T^i_{(0)i},
\end{eqnarray}
and the perturbed field equations in the momentum space
\begin{eqnarray}\label{perturbe eq tt}
8\pi G\delta T^0_0&=&6H\dot{\Phi}-6H^2\Psi+2\frac{\Phi}{a^2}k^2(1+3(\mathbf{b}\cdot\mathbf{\hat{k}}))+\frac{k^2}{2}\Psi(\mathbf{b}\cdot\mathbf{\hat{k}}),\\
\label{perturbe eq ti}
8\pi G\delta T^0_i&=&-i\left(2H\Psi-2\dot{\Phi}\right)k_i,\\
\label{perturbe eq ii}
8\pi G\delta T^i_i/3&=&2\ddot{\Phi}+H\left(6\dot{\Phi}-2\dot{\Psi}\right)-\Psi\left(2H^2+4\frac{\ddot{a}}{a}\right)\nonumber\\
&&+\frac{k^2}{a^2}\left(\Phi+\Psi\right)(1+2(\mathbf{b}\cdot\mathbf{\hat{k}}))+k^2(\mathbf{b}\cdot\mathbf{\hat{k}}) \left(\frac{2}{3a^2}\Phi-\frac{1}{2}\Psi\right),\\
\label{perturbe eq ij}
8\pi G\delta T_{ij}(\hat{k}^i\hat{k}^j-\bar{g}^{ij}/3)&=&\frac{2}{3}k^2(\Phi+\Psi)(1+2(\mathbf{b}\cdot\mathbf{\hat{k}}))+ k^2(\mathbf{b}\cdot\mathbf{\hat{k}})\left(\frac{2}{3}\Phi+1\right).
\end{eqnarray}
The term $\delta T_{ij}(\hat{k}^i\hat{k}^j-\bar{g}^{ij}/3)$ on the left-hand-side of equation (\ref{perturbe eq ij}) denotes the anisotropic stress. The scalar perturbation of the energy-momentum tensor of perfect fluid does not have anisotropic stress \cite{Mukhanov book}, hence $\delta T_{ij}(\hat{k}^i\hat{k}^j-\bar{g}^{ij}/3)=0$. Therefore, equation (\ref{perturbe eq ij}) reduces to
\begin{equation}\label{PsiPhi relation}
-\Psi\simeq\Phi+\frac{3}{2}(\mathbf{b}\cdot\mathbf{\hat{k}}).
\end{equation}
From equation (\ref{PsiPhi relation}) we can see that Finsler spacetime (\ref{FRW like}) will affect the anisotropic part of gravitational field equation.

Here, the energy-momentum tensor in the above field equations is derived by the variation of action (\ref{action}) with respect to the metric. It is of the form
\begin{equation}
T_{\mu\nu}=\partial_\mu\phi\partial_\nu\phi-g_{\mu\nu}\left(\frac{1}{2}g^{\alpha\beta}\partial_\alpha\phi\partial_\beta\phi-V(\phi)\right),
\end{equation}
where $T^0_{(0)0}$ and $T^i_{(0)i}$ in equations (\ref{field eq t},\ref{field eq i}) correspond to the zeroth-order part of inflation field, i.e. $\phi_0(t)$, and $\delta T^\mu_\nu$ in equations (\ref{perturbe eq tt}--\ref{perturbe eq ij}) corresponds to the first-order part of inflation field, i.e. $\delta\phi(t,\vec{x})$. At horizon crossing, by making use of the long wavelength approximation, we can neglect the terms proportional to $k^2$ in equations (\ref{perturbe eq tt}--\ref{perturbe eq ij}). Thus, following the approach of standard inflation model \cite{Mukhanov book}, we find from the field equations (\ref{field eq t}-\ref{perturbe eq ij}) that
\begin{equation}\label{conserved quan}
\left(\mathcal{H}\frac{\delta\phi}{\phi_0'}-\Phi\right)'=0,
\end{equation}
where the prime denotes the derivative with respect to conformal time $\eta\equiv\int\frac{dt}{a}$, and $\mathcal{H}\equiv\frac{a'}{a}$.
Equation (\ref{conserved quan}) implies that the comoving curvature perturbation $R_c\equiv\mathcal{H}\frac{\delta\phi}{\phi_0'}-\Phi$ is conserved outside the Hubble horizon. The comoving curvature perturbation $R_c$ is the same to that in the standard inflation model. This is expected from the qualitative analysis at the beginning of this section. Now, we can use equation (\ref{spectrum delta phi}) to obtain the primordial power spectrum for $R_c$. It is of the form
\begin{equation}\label{spectrum delta Rc}
\mathcal{P}_{R_c}(\mathbf{k})=\mathcal{P}_{iso}(k)\left(1-\frac{15}{2}\mathbf{b}\cdot\hat{\mathbf{k}}\right),
\end{equation}
where $\mathcal{P}_{iso}$ is the isotropic power spectrum for $R_c$.

\section{CMB power spectra with dipolar modulation}\label{sec:spectrum}
The amplitude $\mathbf{b}(k)$ of the dipolar modulation of the primordial power spectrum (\ref{spectrum delta Rc}) can be constrained by the CMB temperature map. Ref.\cite{Aiola} has given the constraint on the scale dependence of the amplitude $A(l)$ of the dipole modulation over the multipole range $2<l<600$. To relate $\mathbf{b}(k)$ to $A(l)$, we note that they have the same effect on the CMB correlation coefficients. In Ref.\cite{Aiola}, the authors expanded the dipolar modulation $\Delta T(\hat{n})=\Delta T_{\rm iso}(\hat{n})(1+A(l)\hat{p}\cdot\hat{n})$ into the usual spherical harmonics, and derived the correlation coefficients of the spherical harmonics, i.e. $C_{XX',ll',mm'}\equiv\langle a^\ast_{l'm'}a_{lm}\rangle$. The off-diagonal correlation corresponds to the dipolar term $A(l)\hat{p}\cdot\hat{n}$. In our paper, we derive the dipolar modulation of primordial power spectrum in Finsler spacetime. The dipolar modulation of primordial power spectrum could also induce the off-diagonal correlation $C_{XX',ll',mm'}$, which will be discussed in the rest of this section. Ref.\cite{Aiola} has constrained $A(l)$ from the Planck date by using the correlation. We could choose the amplitude of the dipolar modulation $\mathbf{b}(k)$ to be the same form as $A(l)$, if we want to account for the correlation. Therefore, the relation between $A(l)$ and $\mathbf{b}(k)$ can be written as
\begin{equation}
|\mathbf{b}(k)|\simeq\frac{4}{15}A(kr_{ls})=\frac{4}{15}A_0\left(\frac{kr_{ls}}{60}\right)^n,
\end{equation}
where we have used the approximate relation $l\sim kr_{ls}$ in deriving the above equation, and the distance to the last scattering surface $r_{ls}$ is approximately $14$ Gpc. The constraint on the scale dependence of the amplitude $A(l)$ in Ref.\cite{Aiola} gives $A_0=0.031^{+0.011}_{-0.012}$ and $n=-0.64^{+0.19}_{-0.14}$. In this paper, we will use the central values of $A_0$ and $n$ given in Ref.\cite{Aiola} to calculate the angular correlation coefficients.

The general angular correlation coefficients that describe the anisotropic effect are given by $C_{XX',ll',mm'}$ \cite{off-diagonal}, where $X$ represents $T$ or $E$. By making use of equation (\ref{spectrum delta Rc}), we obtain the CMB correlation coefficients for scalar perturbations,
\begin{equation}\label{angular correlation}
C_{XX',ll',mm'}=\int\frac{d\ln k}{(2\pi)^3}\Delta_{X,l0}(k)\Delta^\ast_{X',l'0}(k)P_{ll'mm'},
\end{equation}
where
\begin{eqnarray}
P_{ll'mm'}&=&\int d\Omega\mathcal{P}_{R_c}Y^\ast_{lm}Y_{l'm'}\nonumber\\
\label{angular correlation1}
&=&\mathcal{P}_{iso}\delta_{mm'}\left(\delta_{ll'}-2A_0\left(\frac{kr_{ls}}{60}\right)^n\sqrt{\frac{2l'+1}{2l+1}} \mathcal{C}^{l'm}_{10lm}\mathcal{C}^{l'0}_{10l0}\right),
\end{eqnarray}
$\Delta_{X,ls}(k)$ denote the transfer functions, and $\mathcal{C}^{lm}_{LMl'm'}$ are the Clebsch-Gordan coefficients. The CMB correlation coefficients $C_{ll'}\equiv C_{XX',ll',mm'}$ contain two parts. The one proportional to $\delta_{ll'}$ denotes the statistical isotropy of the CMB. The other proportional to $\mathcal{C}^{l'm}_{10lm}\mathcal{C}^{l'0}_{10l0}$ denotes the statistical anisotropy of the CMB. The Clebsch-Gordan coefficients do not vanish only if $l'=l\pm1$. This property implies that our Finslerian modification of primordial power spectrum does not affect the isotropic part of correlation coefficients $C_{ll'}$ which describes the statistical isotropy of the CMB. This is different from the anisotropic inflation models that induce quadruple modulation of the primordial power spectrum \cite{off-diagonal,Komatsu limit}.

Ref.\cite{Aiola} has used the Planck 2013 data to constrain the amplitude $A(l)$ of the dipole modulation. To be consistent with their results, we adopt the best-fitting central value of cosmological parameters released by the Planck 2013 data and use the formula of angular correlation coefficients (\ref{angular correlation},\ref{angular correlation1}) to plot numerical results for the anisotropic part of $C_{ll'}$. One can find from equations (\ref{angular correlation},\ref{angular correlation1}) that the anisotropic part of $C_{ll'}$ depends on $m$, and the $TE$ and $ET$ correlation coefficients are different.

\begin{figure}
\includegraphics[width=0.49\textwidth]{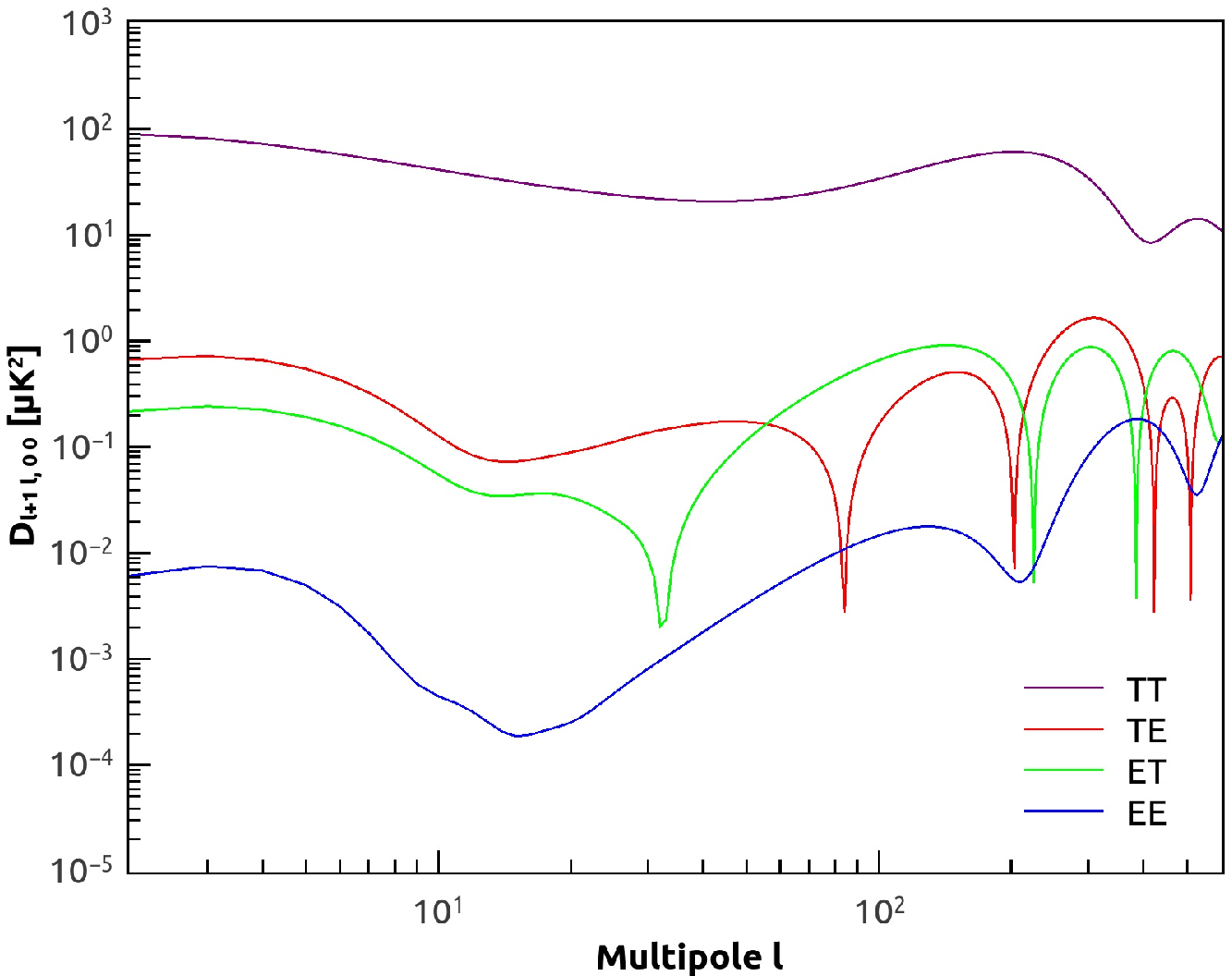}
\includegraphics[width=0.49\textwidth]{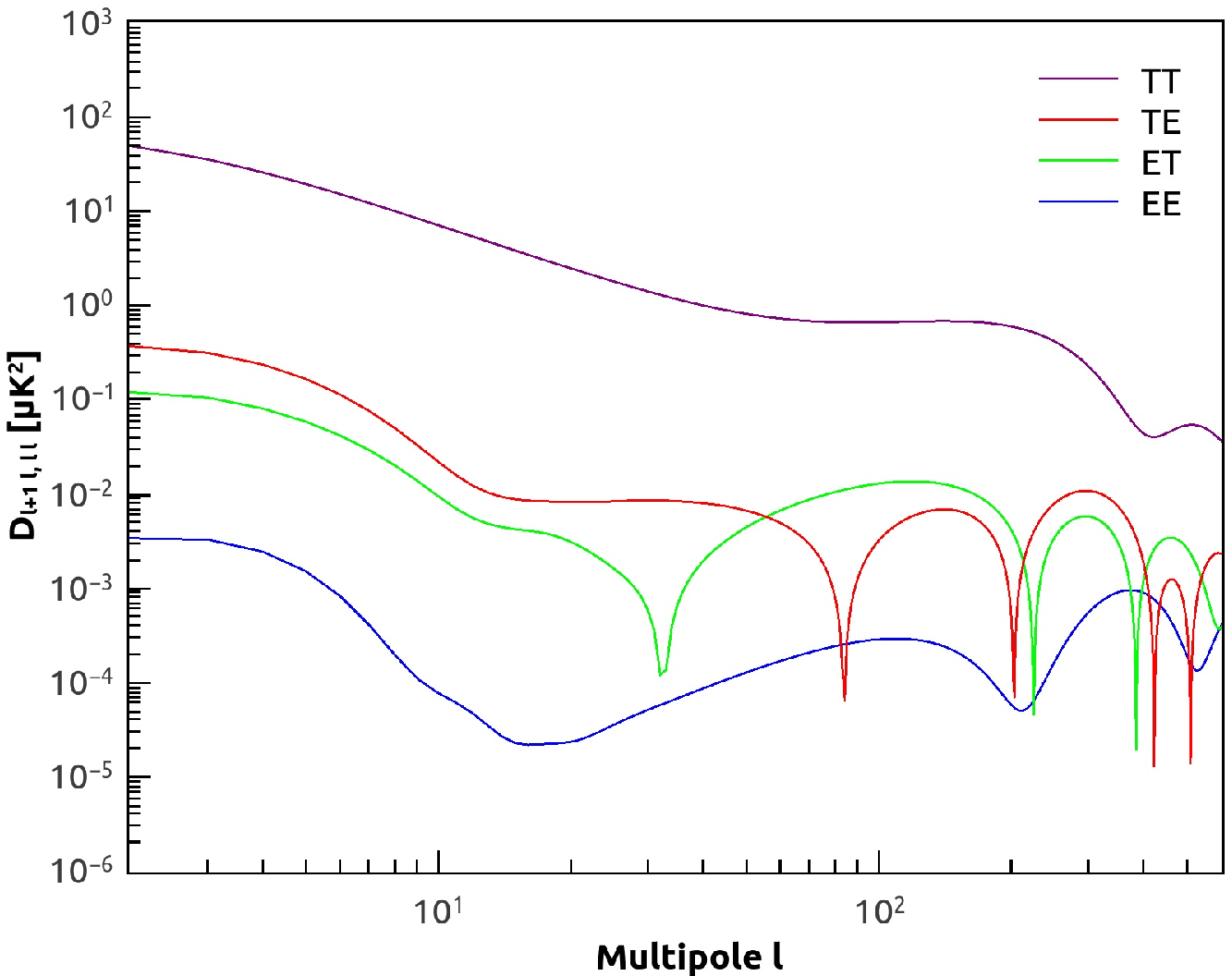}
\caption{The anisotropic $TT$, $TE$, $ET$, $EE$ correlation coefficients for $m=0$ (left panel) and $m=l$ (right panel).}
\label{anisotropic correlation}
\end{figure}

The anisotropic $TT$, $TE$, $ET$, $EE$ correlation coefficients $D^{XX'}_{ll',mm'}$ are plotted in Fig.\ref{anisotropic correlation}. Here, the coefficients $D^{XX'}_{ll',mm'}$ are defined as
\begin{equation}
D^{XX'}_{ll',mm'}\equiv (2\pi)^{-1}\sqrt{l(l+1)l'(l'+1)}\,|C_{XX',ll',mm'}|.
\end{equation}
The correlation coefficients in $m=0$ and $m=l$ cases are plotted in the left panel and right panel, respectively.
From this figure we can see that the amplitude of anisotropic coefficients for $m=l$ is relatively smaller than that for $m=0$. This is due to fact that we have already chosen the dipolar direction along $z$-axis.

\section{Conclusions and Remarks}\label{sec:conclusion}

In this paper, we investigated the dipolar modulation of CMB power spectrum in the framework of Finsler spacetime. Differing from most previous models in which the primordial power spectrum $P(k,\mathbf{r})$ is spatially varying, the model we derived here allows a three dimensional spectrum $P(\mathbf{k})$ with a preferred direction. This is due to the fact that the Randers spacetime, a special kind of Finsler spacetime, is non-reversible under parity flip, $\mathbf{x}\rightarrow-\mathbf{x}$. Such a property makes the Fourier component $\delta(\mathbf{k})$ does not satisfy the relation $\delta(-\mathbf{k})=\delta^\ast(\mathbf{k})$. Thus, a three dimensional spectrum $P(\mathbf{k})$ with preferred direction appears if the background spacetime is Randers spacetime during the stage of inflation. Following the similar approach of standard inflation model \cite{Mukhanov book}, we found from the gravitational field equations that the comoving curvature perturbation $R_c$ is the same to the one in the standard inflation model. Then, we used the primordial power spectrum of $R_c$ to derive the general angular correlation coefficients $C_{XX',ll',mm'}$. The correlation coefficients contain two parts. One represents the statistical isotropy of the CMB, which is similar to that in the $\Lambda$CDM model. The other represents the statistical anisotropy of the CMB, which does not vanish only if $l'=l\pm 1$. This is the main feature of the dipolar modulation. Finally, we used the mean value of the amplitude of the dipole modulation in the multipole range $2 < l < 600$ given by Ref.\cite{Aiola} to obtain the numerical results for the anisotropic part of correlation coefficients.

\begin{acknowledgments}
We thank Prof. Z. Chang and Dr. S. Wang for helpful discussions. This work has been supported by the National Natural Science Fund of China (Grant Nos. 11305181, 11603005 and 11647307), the Open Project Program of State Key Laboratory of Theoretical Physics, Institute of Theoretical Physics, Chinese Academy of Sciences, China (Gront No. Y5KF181CJ1), and the Fundamental Research Funds for the Central Universities (Grant No. 106112016CDJCR301206).
\end{acknowledgments}


\begin{thebibliography}{999}
\bibitem{Sahni}V. Sahni, Class. Quant. Grav. {\bf 19}, 3435 (2002).
\bibitem{Padmanabhan}T. Padmanabhan, Phys. Rept. {\bf 380}, 235 (2003).
\bibitem{Starobinsky}A. A. Starobinsky, Phys. Lett. B {\bf 91}, 99 (1980).
\bibitem{Sato:1981}K. Sato, Mon. Not. Roy. Astron. Soc. {\bf 195}, 467 (1981).
\bibitem{Guth:1981}A. H. Guth, Phys. Rev. D {\bf 23}, 347 (1981).
\bibitem{Linde:1982}A. D. Linde, Phys. Lett. B {\bf 108}, 389 (1982).
\bibitem{Albrecht:1982}A. Albrecht and P. J. Steinhardt, Phys. Rev. Lett. {\bf 48}, 1220 (1982).
\bibitem{Riotto}A. Riotto, arXiv:hep-ph/0210162.
\bibitem{Ade:2013sjv}P.~A.~R.~Ade {\it et al.} [Planck Collaboration], Astron.\ Astrophys.\ {\bf 571}, A1 (2014).
\bibitem{Adam:2015rua}R.~Adam {\it et al.} [Planck Collaboration], Astron.\ Astrophys.\  {\bf 594}, A1 (2016).
\bibitem{Power asymmetry1}P.~A.~R.~Ade {\it et al.} [Planck Collaboration], Astron. Astrophys. {\bf 571}, A23 (2014).
\bibitem{Power asymmetry2}P.~A.~R.~Ade {\it et al.} [Planck Collaboration], Astron. Astrophys. {\bf 594}, A16 (2016).
\bibitem{Akrami:2014eta}Y.~Akrami, Y.~Fantaye, A.~Shafieloo, H.~K.~Eriksen, F.~K.~Hansen, A.~J.~Banday and K.~M.~G¨®rski, Astrophys.\ J.\  {\bf 784}, L42 (2014).
\bibitem{Eriksen:2003db}H.~K.~Eriksen, F.~K.~Hansen, A.~J.~Banday, K.~M.~Gorski and P.~B.~Lilje, Astrophys.\ J.\  {\bf 605}, 14 (2004) [Erratum: Astrophys.\ J.\  {\bf 609} (2004) 1198]
\bibitem{Hansen:2004vq}F.~K.~Hansen, A.~J.~Banday and K.~M.~Gorski, Mon.\ Not.\ Roy.\ Astron.\ Soc.\  {\bf 354}, 641 (2004).
\bibitem{Eriksen:2007pc}H.~K.~Eriksen, A.~J.~Banday, K.~M.~Gorski, F.~K.~Hansen and P.~B.~Lilje, Astrophys.\ J.\  {\bf 660}, L81 (2007).
\bibitem{Hansen:2008ym}F.~K.~Hansen, A.~J.~Banday, K.~M.~Gorski, H.~K.~Eriksen and P.~B.~Lilje, Astrophys.\ J.\  {\bf 704}, 1448 (2009).
\bibitem{Hoftuft:2009ym}J. Hoftuft, H. K. Eriksen, A. J. Banday, K. M. Gorski, F. K. Hansen, P. B. Lilje, Astrophys. J. {\bf 699}, 985 (2009).
\bibitem{Gordon}C. Gordon, W. Hu, D. Huterer, and T. Crawford, Phys. Rev. D {\bf 72}, 103002 (2005).
\bibitem{Dai:2013}L. Dai, D. Jeong, M. Kamionkowski and J. Chluba, Phys. Rev. D {\bf 87}, 123005 (2013).
\bibitem{Lyth:2013}D. H. Lyth, JCAP {\bf 1308}, 007(2013).
\bibitem{Pullen}A. R. Pullen and M. Kamionkowski, Phys. Rev. D {\bf 76}, 103529 (2007).
\bibitem{Erickcek:2008sm}A.~L.~Erickcek, M.~Kamionkowski and S.~M.~Carroll, Phys.\ Rev.\ D {\bf 78}, 123520 (2008).
\bibitem{Lyth:2013vha}D.~H.~Lyth, JCAP {\bf 1308}, 007 (2013).
\bibitem{McDonald:2013qca}J.~McDonald, JCAP {\bf 1311}, 041 (2013).
\bibitem{Jazayeri:2014nya}S.~Jazayeri, Y.~Akrami, H.~Firouzjahi, A.~R.~Solomon and Y.~Wang, JCAP {\bf 1411}, 044 (2014).
\bibitem{Assadullahi:2014pya}H.~Assadullahi, H.~Firouzjahi, M.~H.~Namjoo and D.~Wands, JCAP {\bf 1504}, 017 (2015).
\bibitem{Kenton:2015jga}Z.~Kenton, D.~J.~Mulryne and S.~Thomas, Phys.\ Rev.\ D {\bf 92}, 023505 (2015).
\bibitem{Byrnes:2015dub}C.~T.~Byrnes, D.~Regan, D.~Seery and E.~R.~M.~Tarrant, JCAP {\bf 1606}, 025 (2016).
\bibitem{Byrnes:2016uqw}C.~Byrnes, G.~Dom¨¨nech, M.~Sasaki and T.~Takahashi, JCAP {\bf 1612}, 020 (2016).
\bibitem{Bianchi spacetime}K. Rosquist, R. T. Jantzen, Phys. Rep. {\bf 166}, 89 (1988).
\bibitem{Koivisto:2008}T. Koivisto and D. F. Mota, JCAP {\bf 0806}, 018 (2008).
\bibitem{Koivisto:2011}L. Campanelli, P. Cea, G. L. Fogli and A. Marrone, Phys. Rev. D {\bf 83}, 103503 (2011).
\bibitem{Randers}G. Randers, Phys. Rev. {\bf 59}, 195 (1941).
\bibitem{Book by Bao}D. Bao, S. S. Chern, and Z. Shen, {\it An Introduction to Riemann--Finsler Geometry}, Graduate Texts in Mathematics {\bf 200}, Springer, New York, 2000.
\bibitem{Finsler dipolar modulation}X. Li, S. Wang and Z. Chang, Eur. Phys. J. C {\bf 75}, 260 (2015).
\bibitem{Hanson}D. Hanson and A. Lewis, Phys. Rev. D {\bf 80}, 063004 (2009).
\bibitem{Ade:2013nlj}P.~A.~R.~Ade {\it et al.} [Planck Collaboration], Astron.\ Astrophys.\  {\bf 571}, A23 (2014).
\bibitem{Flender:2013jja}S.~Flender and S.~Hotchkiss, JCAP {\bf 1309}, 033 (2013).
\bibitem{Ade:2015hxq}P.~A.~R.~Ade {\it et al.} [Planck Collaboration], Astron.\ Astrophys.\  {\bf 594}, A16 (2016).
\bibitem{Aiola}S. Aiola, B. Wang, A. Kosowsky, T. Kahniashvili and H. Firouzjahi, Phys. Rev. D. {\bf 92}, 063008 (2015).
\bibitem{Chang:2013lfm}Z. Chang, M.-H. Li, X. Li and S. Wang, Eur. Phys. J. C {\bf 73}, 2459 (2013).
\bibitem{Chang:2014ndh}Z. Chang, X. Li, H.-N. Lin and S. Wang, Eur. Phys. J. C {\bf 74}, 2821 (2014).
\bibitem{Chang:2014smk}Z. Chang, X. Li, H.-N. Lin and S. Wang, Mod. Phys. Lett. A {\bf 29}, 1450067 (2014).
\bibitem{Chang:2015nbc}X. Li, H.-N. Lin, S. Wang and Z. Chang, Eur. Phys. J. C {\bf 75}, 181 (2015).
\bibitem{Rev inflaton1}A. Mazumdar, J. Rocher, Phys. Rept. {\bf 497}, 85 (2011).
\bibitem{Rev inflaton2}L. Wang, E. Pukartas, A. Mazumdar, JCAP {\bf 07}, 019 (2013).
\bibitem{Finsler tensor spectrum}X. Li and S. Wang, Eur. Phys. J. C {\bf 76}, 51 (2016).
\bibitem{Mukhanov book}V. Mukhanov, {\it Physical Foundations of Cosmology}, Cambrdige Uni. Press, 2005.
\bibitem{Finsler bullet}X. Li, M.-H. Li, H.-N. Lin, and Z. Chang, Mon. Not. R. Astron. Soc. {\bf 428}, 2939 (2013).
\bibitem{Finsler BH}X. Li and Z. Chang, Phys. Rev. D {\bf90}, 064049 (2014).
\bibitem{Akbar}H. Akbar-Zadeh, Acad. Roy. Belg. Bull. Cl. Sci. {\bf 74}, 281 (1988).
\bibitem{off-diagonal}M. Watanabe, S. Kanno, and J. Soda, Mon. Not. R. Astron. Soc. {\bf 412}, L83 (2011).
\bibitem{Komatsu limit}J. Kim and E. Komatsu, Phys. Rev. D {\bf 88}, 101301(R) (2013).



\end{thebibliography}
\end{document}